\documentclass[twocolumn, secnumaRabic, amssymb, nobibnotes, aps, showpacs,superscriptaddress]{revtex4-1}
\usepackage{graphicx,subfigure,amsmath}

\usepackage{amsmath,amstext}
\usepackage{graphicx}
\usepackage{booktabs}
\usepackage{multirow}
\usepackage{subfigure}
\usepackage{dcolumn}
\usepackage{mathrsfs}
\usepackage{color}
\usepackage{multirow}
\usepackage{afterpage}
\usepackage{subfigure}

\begin{document}
	\title{Gain on ground state of quantum system for truly $\mathcal{PT}$ symmetry}
	\author{Bing-Bing Liu}
	\affiliation{School of Physics, Zhengzhou University, Zhengzhou 450001, China}
	\author{Shi-Lei Su}
	\email{slsu@zzu.edu.cn}
	\affiliation{School of Physics, Zhengzhou University, Zhengzhou 450001, China}
	\affiliation{Institute of Quantum Materials and Physics, Henan Academy of Sciences, Zhengzhou 450046, China}
	\affiliation{Institute of Quantum Science and Technology, Yanbian University, Yanji, Jilin 133002, China}
	\date{\today}	
	\begin{abstract}
		For a truly $\mathcal{PT}$-symmetric quantum system, the conventional non-Hermitian Hamiltonian is $H = \Omega\sigma_x -i\gamma|1\rangle\langle1| + i\gamma|0\rangle\langle0|$, where $\Omega$ and $\gamma$ are real parameters and $\sigma_x$ denotes Pauli X operator. These three terms represent coherent coupling, loss (on state $|1\rangle$), and gain (on state $|0\rangle$), respectively. Although the works in [Phys. Rev. Lett. \textbf{101}, 230404 (2008); Phys. Rev. Lett. \textbf{119}, 190401 (2017); Science \textbf{364}, 878 (2019)] proposed theoretically and/or demonstrate dilation methods for a truly parity-time($\mathcal{PT}$)–symmetric Hamiltonian by embedding into larger Hermitian space, directly realizing the gain term $+i\gamma|0\rangle\langle0|$ has still remained an outstanding challenge for quantum system. While systems omitting this gain term can exhibit a passively $\mathcal{PT}$-symmetric energy spectrum (featuring a parallel imaginary shift) and display related phenomena, they fail to capture the full physical behavior and unique properties inherent to truly $\mathcal{PT}$-symmetric systems. In this manuscript, we propose a method to achieve effective gain on the ground state $|0\rangle$ ($+i\gamma|0\rangle\langle0|$) after averaging all trajectories, by integrating the S{\o}rensen-Reiter effective operator method with the Wiseman-Milburn master equation for continuous measurement and instantaneous feedback control after averaging the evolution over all trajectories. This approach provides a possible pathway to efficiently construct truly $\mathcal{PT}$-symmetric quantum devices, offering a powerful platform for engineering quantum resources vital for quantum information technology applications.
	\end{abstract}

	\maketitle
	
	\section{Introduction}

	Non-Hermitian parity-time ($\mathcal{PT}$)-symmetry, introduced by Bender and Boettcher in 1998~\cite{PhysRevLett.80.5243,RevModPhys.96.045002}, extends the traditional framework of quantum mechanics by demonstrating that certain non-Hermitian Hamiltonians can possess real energy spectra. A Hamiltonian \( H \) is $\mathcal{PT}$-symmetric if it commutes with the combined parity-time reversal operator, i.e., \( [H, \mathcal{PT}] = 0 \), where \( \mathcal{P} \) reverses spatial coordinates and \( \mathcal{T} \) reverses time and complex-conjugates wavefunctions. For such Hamiltonians, the eigenvalues can be entirely real when the $\mathcal{PT}$-symmetry is unbroken, which typically requires that the potential has even real parts and odd imaginary parts in space. This corresponds to a balanced distribution of gain and loss in the system. This delicate balance gives rise to unique phenomena such as exceptional points (EPs)~\cite{Moiseyev2011,Bagarello2016,Ashida02072020,PhysRevLett.127.186601,PhysRevLett.128.249901,PhysRevLett.127.186602}, where eigenvalues and eigenvectors coalesce, and non-Hermitian topological phases~\cite{PhysRevLett.118.040401,PhysRevLett.120.146402,PhysRevX.8.031079,PhysRevX.9.041015,PhysRevB.98.085116}, which are distinct from their Hermitian counterparts and emerge from the dynamic equilibrium in open quantum systems.

	Active  $\mathcal{PT}$-symmetry (balanced  gain and loss) has been successfully realized in various classical wave systems, particularly in optics, photonics and acoustics~\cite{feng2014,Jahromi2017,Chang2014,Ruter2010,Liu2018,Sahin2019,miri2019,feng2017,zhao2018,li2020,zhang2020,zhu2014,shi2016}, electrical circuit~\cite{Assawaworrarit2017}, excitation-polariton~\cite{Gao2015}, and lattice~\cite{Regensburger2012,Kremer2019,Zhang2016}, through the careful engineering of gain and loss distributions. In photonic systems, coupled waveguides or resonators are designed with spatially separated regions of gain (achieved via optical pumping) and loss (using absorptive materials), creating synthetic $\mathcal{PT}$-symmetric lattices. Notable examples include fiber ring resonators with electro-optic modulation, which have demonstrated unidirectional invisibility and coherent perfect absorption \cite{Ruter2010}. In acoustics, $\mathcal{PT}$-symmetric systems have been realized using flow duct acoustics, where the mean flow induces effective gain and loss, leading to  $\mathcal{PT}$-symmetric scattering \cite{Auregan2017}. The rest systems are the same to get active $\mathcal{PT}$ by balancing gain and loss. Additionally, passive  $\mathcal{PT}$-symmetric acoustic systems have been demonstrated using metamaterials with carefully designed unit cells that modulate both the real and imaginary parts of the refractive index, resulting in asymmetric diffraction \cite{Li2019}. 
	However, the energy spectra of passive $\mathcal{PT}$-symmetry often includes imaginary components due to net loss, which can lead to exponential decay of wave functions unless mitigated by specific designs. 
	
	While numerous schemes have explored $\mathcal{PT}$-symmetry in quantum atomic systems, these predominantly involve passive configurations, leveraging naturally occurring loss mechanisms like spontaneous decay from excited states~\cite{2023speed}. Achieving active gain on the ground state, essential for realizing balanced gain and loss on atomic energy levels, remains a fundamental challenge that has yet to be demonstrated. Although recent work \cite{Liu2023} demonstrates active $\mathcal{PT}$-symmetry in the collective atomic spin phase space using real-time feedback control, it does not achieve the balanced gain and loss on the atomic levels that is the focus of our investigation. Besides, the inspiring works~\cite{Uwe2008,Kohei2017,djf2019} use the dilation method to achieve the $\mathcal{PT}$-symmetry by embedding into larger Hermitian space, which may require increased system complexity/dimensions for scalability.
	
	In this manuscript, we address the challenge of engineering gain on ground state of quantum system by integrating the Wiseman–Milburn quantum feedback  (WMQF) theory~\cite{wiseman1993,Guilmin2023,Nelson2000,wiseman1994,wm1994,Zhang2017,wisemanbook} with the Sørensen–Reiter effective operator method~\cite{Kastoryano2011,Reiter2012,ReiterNJP} (SEOM). WMQF theory provides a robust framework for modeling continuously monitored quantum systems, where real-time homodyne measurement generates a stochastic current that can be fed back into system operators to steer dynamics beyond passive evolution. Complementarily, SEOM employs auxiliary states to systematically eliminate rapidly decaying degrees of freedom, enabling precise modulation of the dissipation operator’s form and decay strength. By combining these methods, we achieve a hybrid strategy that synthesizes autonomous gain in the ground state, offering a versatile approach to tailoring quantum dynamics in open systems.
	
	The paper are ognized as follows: In Sec.~\ref{sec2}, we review the basic process of WMQF theory and S{\o}rensen-Reitor effective operator method, respectively. In Sec.~\ref{sec3}, we construct the dynamics to achieve the gain on ground state with an effective way. In Sec.~\ref{sec4}, we give the discussion and the numerical verifications of the scheme and conclusion is given in  Sec.~\ref{sec5}.  
	
	\section{Review of WMQF and SEOM}\label{sec2} 
	
	\subsection{Wiseman-Milburn Feedback control}
	
	\begin{figure}[h]
		\centering
		\includegraphics[width=\linewidth]{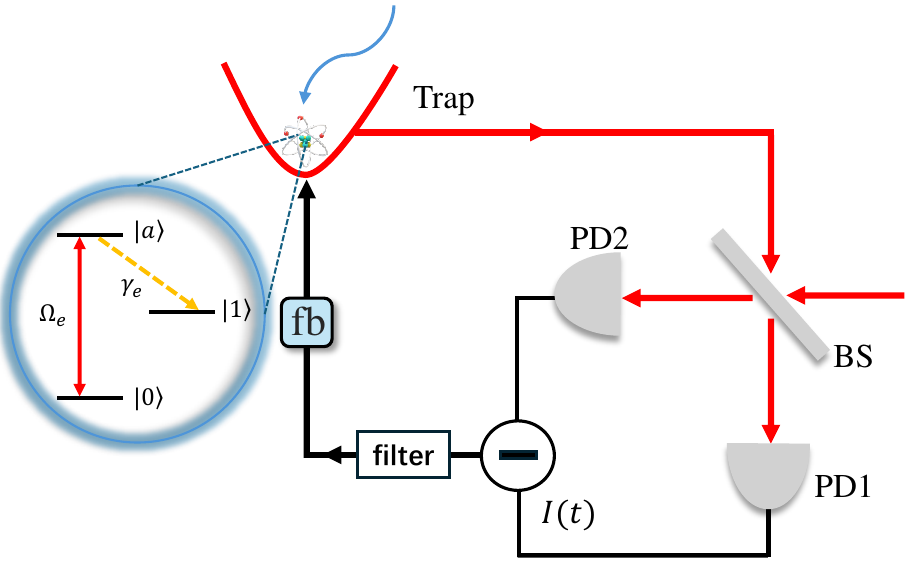}
		\caption{
			Schematic of the Wiseman-Milburn continuous feedback protocol. The quantum system contains three levels with $|0\rangle$, $|1\rangle$ and $|a\rangle$ being ground state, excited state and auxiliary state, respectively. The system undergoes continuous monitoring via homodyne detection, characterized by the effective measurement operator \( L_{\text{eff}} = \Gamma_{\text{eff}} |1\rangle\langle 0| \), derived using the Reitor effective operator method. The feedback operator, defined as \( F = |0\rangle\langle 1| + |1\rangle\langle 0| \), is Hermitian and facilitates the feedback mechanism. The feedback strength \( G \), chosen to be negative for gain conversion, modulates the feedback Hamiltonian. The photocurrent increment is expressed as \( I(t)dt = \sqrt{\gamma_1} \langle F \rangle_c dt + dW \), where the subscript \( c \) indicates conditional dynamics dependent on the measurement record. PD$_j$ is photondetector with $j = 1,2$.
		}
		\label{fig1}
	\end{figure}

	In 1993, Wiseman and Milburn extended quantum-trajectory formalism to mixed states by deriving a stochastic master equation (SME) that incorporates both the system’s intrinsic dynamics and the stochastic back-action of inefficient continuous measurements~\cite{wiseman1993}.  This SME makes it possible to track the observer’s state-of-knowledge when the detector has non-unit efficiency and the measurement record is noisy~\cite{Guilmin2023}, a situation that is common in experiments with trapped atoms~\cite{Nelson2000}. On that basis, Wiseman presented a general theory of quantum-limited feedback for continuously monitored systems and showed that if the measurement current is fed back instantaneously into a Hamiltonian control term, with the schematic diagram shown in Fig.~\ref{fig1}, the unconditional evolution of the system obeys a Markovian master equation~\cite{wiseman1994,wm1994,wisemanbook,Zhang2017}. And this method has been used  to stabilize arbitrary one-qubit quantum states~\cite{Wang115,Hofmann116,Hofmann117}, manipulate quantum entanglement~\cite{Wang118,Li119,Yamamoto120,Mancini121,Li122,Carvalho123,Viviescas124,Carvalho125,Carvalho126,Hou127,Stevenson128,Wang129,Xue130,Song131}, generate and protect Schrödinger cat states~\cite{Tombesi132,Goetsch133,Vitali134,Vitali135,Fortunato136}, and induce optical, mechanical, and spin squeezing~\cite{Wiseman137,Wiseman138,Liebman139,Tombesi140,Wiseman141,Wiseman142,Ruskov143,Vinante144,Thomsen145}.

	We first introduce the concept of photocurrent that is basic for WMQF, which is essential for feedback control. It is defined as
	\begin{equation}
		I(t)dt = \sqrt{\gamma_1} \langle F \rangle_c dt + dW,
	\end{equation}
	where \( \sqrt{\gamma_1} \langle F \rangle_c dt \) represents the signal proportional to the system observable \( F \), with \( \langle F \rangle_c = \text{Tr}(F \rho_c) \) being the conditional expectation based on prior measurement outcomes. And \( dW \) is the Wiener increment, encapsulating quantum shot noise, with statistical properties \( \mathbb{E}[dW] = 0 \) and \( \mathbb{E}[(dW)^2] = dt \)~\cite{wiseman1993,wisemanbook}. The subscript \( c \) denotes conditional dynamics, indicating the system's state evolution based on the measurement record. For example, \( \rho_c(t) \), the conditional density matrix, follows a SME that captures individual quantum trajectories influenced by measurement outcomes. Averaging these trajectories yields the unconditional state: \( \rho(t) = \mathbb{E}[\rho_c(t)] \). Additionally, the conditional expectation \( \langle F \rangle_c = \text{Tr}(F \rho_c) \) provides real-time system insights, essential for feedback control.

	The feedback Hamiltonian is~\cite{wisemanbook}
	\begin{equation}
		H_{\text{fb}}(t) = G \cdot I(t) \cdot F, \quad G \in \mathbb{R},
	\end{equation}
	where \( G \) is the feedback strength. And \( I(t) \) is the photocurrent, linking measurement outcomes to control actions.

	The SME describes the conditional evolution of \( \rho_c \) under continuous measurement without feedback is~\cite{wisemanbook}
	\begin{equation}
		d\rho_c = -i [H_{\text{sys}}, \rho_c] dt + \gamma_1 \mathcal{D}[L] \rho_c dt + \sqrt{\gamma_1} \mathcal{H}[L] \rho_c dW,
	\end{equation} where  \( \mathcal{D}[L] \rho_c = L \rho_c L^\dagger - \frac{1}{2} \{ L^\dagger L, \rho_c \} \) is the dissipation superoperator. And \( \mathcal{H}[L] \rho_c = (L \rho_c + \rho_c L^\dagger) - \langle L + L^\dagger \rangle_c \rho_c \) represents measurement backaction. This equation arises from homodyne detection, combining unitary evolution, dissipation, and stochastic measurement effects.
	
	The feedback operator is
	$ U_{\text{fb}} = e^{-i H_{\text{fb}} dt} = e^{-i G F dy}$
	with $dy = \sqrt{\gamma_1} \langle F \rangle_c dt + dW$. 
	The feedback-modified density matrix becomes
	\begin{equation}
		\rho_c^{\text{fb}} = U_{\text{fb}} (\rho_c + d\rho_c) U_{\text{fb}}^\dagger.
	\end{equation}
	Incorporating the SME and \( dy \), one can get
	\begin{eqnarray}\label{SME}
		\rho_c^{\text{fb}} = && \rho_c - i [H_{\text{sys}}, \rho_c] dt + \gamma_1 \mathcal{D}[L] \rho_c dt + \sqrt{\gamma_1} \mathcal{H}[L] \rho_c dW
		\cr\cr&& - i G [F, \rho_c] (\sqrt{\gamma_1} \langle F \rangle_c dt + dW) + G^2 \mathcal{D}[F] \rho_c dt.
	\end{eqnarray}

	The unconditional density matrix \( \rho(t) = \mathbb{E}[\rho_c(t)] \) averages over noise, with evolution:
	\begin{equation}
		\dot{\rho} = \mathbb{E}\left[ \frac{\rho_c^{\text{fb}} - \rho_c}{dt} \right].
	\end{equation}	
	Using Wiener process properties and the mean-field approximation (\( \mathbb{E}[\langle F \rangle_c \rho_c] \approx \langle F \rangle \rho \)). The resulting unconditional master equation becomes
	\begin{eqnarray}\label{unconditionalmaster}
		\dot{\rho} =&& -i [H_{\text{sys}}, \rho] + \gamma_1 \mathcal{D}[L] \rho 
		- i G \sqrt{\gamma_1} [F, L\rho + \rho L^\dagger] \cr\cr&&+ G^2 \mathcal{D}[F] \rho.
	\end{eqnarray}

	\subsection{Sørensen–Reiter Effective‑Operator Method}
	Assuming the open system consist of two distinct subspaces, one for the ground states and one for the decaying excited states. 
	The couplings of these two subspaces are assumed to be perturbative. Ground state is denoted by $|g\rangle$ and excited state by $|e\rangle$.  Coherent drives denote by $\hat V_+$ and $\hat V_-$. Excited(Ground)‑state Hamiltonian is denoted by $\hat H_{e(g)}$, Dissipative decay $\{\hat L_k\}$  from $P_e \to P_g$.
	\begin{equation}
		P_g = \sum |g\rangle \langle g|,~P_e = \sum |e\rangle \langle e|,~
		P_g + P_e = I,~P_g P_e = 0.
	\end{equation}
	The Hamiltonian and jump operators can be splitted as
	\begin{equation*}
		\hat H = \hat H_g + \hat H_e + \hat V_+ + \hat V_-, 
	\end{equation*} with $\hat H_g = P_g \hat H P_g$,~$\hat H_e = P_e \hat H P_e$, $\hat V_+ = P_e \hat H P_g$, $\hat V_- = P_g \hat H P_e$
	Assuming that $\hat H_e$ and $\{\hat L_k\}$ generate fast dynamics compared to the weak coupling $\hat V = \hat V_+ + \hat V_-$. Define the excited‑manifold Liouvillian
	\begin{equation}
		\mathcal{L}_e(\rho) = -i[\hat H_e, \rho] + \sum_k \mathcal{D}[\hat L_k](\rho),
	\end{equation} 
	with
	$\mathcal{D}[A](\rho) = A \rho A^\dagger - \frac{1}{2} \{A^\dagger A, \rho\}$
	
	Then to second order in $\hat V$, the projected master equation is
	\begin{equation}
		\dot \rho_g \approx P_g \mathcal{L} P_g \rho 
		- P_g \mathcal{V} P_e \mathcal{L}_e^{-1} P_e \mathcal{V} P_g \rho.
	\end{equation}
	After simplification, the effective dynamics on the ground subspace becomes~\cite{Reiter2012,ReiterNJP,Kastoryano2011}
	\begin{equation}
		\dot \rho_g = -i[\hat H_{\text{eff}}, \rho_g] + \sum_k \mathcal{D}[\hat L_{\text{eff}}^{(k)}] \rho_g,
	\end{equation}
	with
	\begin{eqnarray}
		\hat H_{\text{NH}} &=& \hat H_e - \frac{i}{2} \sum_k \hat L_k^\dagger \hat L_k,\\
		\hat H_{\text{eff}} &=& \hat H_g - \frac{1}{2} \hat V_- \left( \hat H_{\text{NH}}^{-1} + \hat H_{\text{NH}}^{-\dagger} \right) \hat V_+,\\
		\hat L_{\text{eff}}^{(k)} &=& \hat L_k \hat H_{\text{NH}}^{-1} \hat V_+.
	\end{eqnarray}

	\section{Non-Hermitian Hamiltonian with gain}\label{sec3}
	
	\subsection{Invalid with natural decay channel}
	We start from the unconditional feedback master equation
	\begin{equation}
		\dot{\rho}
		= -\,i\bigl[H_{\rm sys},\rho\bigr]
		+ \gamma_1\,\mathcal{D}[L]\,\rho
		- i\,G\sqrt{\gamma_1}\,\bigl[F,\,L\,\rho + \rho\,L^\dagger\bigr]
		+ G^2\,\mathcal{D}[F]\,\rho,
		\label{eq:origME}
	\end{equation}
	where
	$
	\mathcal{D}[A]\rho
	= A\,\rho\,A^\dagger
	- \tfrac{1}{2}\bigl\{A^\dagger A,\rho\bigr\}$, and $F = F^\dagger$ is satisfied.
	
	One verifies by direct expansion that the sum of dissipators and commutators in Eq.~\eqref{eq:origME} can be written as a single Lindblad term,
	\begin{equation}
		\gamma_1\,\mathcal{D}[L]
		- i\,G\sqrt{\gamma_1}\,[F,\,L\,\rho + \rho\,L^\dagger]
		+ G^2\,\mathcal{D}[F]
		\;=\;
		\mathcal{D}\bigl(\tilde{c}\bigr)\,\rho,
	\end{equation}
	provided we define the \emph{effective collapse operator}
	\begin{equation}
		\tilde{c}
		= \sqrt{\gamma_1}\,L \;-\; i\,G\,F.
	\end{equation}
	Hence the master equation becomes
	\begin{equation}
		\dot{\rho}
		= -\,i\bigl(H_{\rm eff}\,\rho - \rho\,H_{\rm eff}^\dagger\bigr)
		\;+\;\tilde c\,\rho\,\tilde c^\dagger.
	\end{equation}
	with	$H_{\rm eff}
	= H_{\rm sys}
	\;-\;\frac{i}{2}\,\tilde c^\dagger \tilde c$
	provided $L^{\dagger}F=-FL$.
	Substituting $\tilde c$ into \(H_{\rm eff}\) gives
	\begin{equation}\label{Hnh}
		H_{\rm eff}
		= H_{\rm sys}
		- \frac{i\gamma_1}{2}\,L^\dagger L
		- \frac{G\sqrt{\gamma_1}}{2}\,\bigl[L^\dagger F - F\,L\bigr].
	\end{equation}
	Since \(F^2\propto\mathbb{I}\) contributes only a global energy shift, the corresponding term is dropped.  
	
	By defining the single jump operator \(\tilde{c}\), one chooses a specific unraveling into “drift + jumps.”
	Interpret that drift as generating the no‑jump evolution in that jump unraveling. Dropping the jump term is then fully justified whenever either (i) post‑select on trajectories with zero detection clicks, or (ii) restrict attention to time intervals so short (or jump rates so low) that the probability of any jump is negligible. Dropping the jump term \(\tilde{c}\rho\,\tilde{c}^\dagger\) isolates the \emph{conditional} (no‑jump) evolution, either via post‑selection on no clicks or in the short‑time/weak‑jump limit \(\langle \tilde{c}^\dagger \tilde{c}\rangle \,\Delta t \ll 1\).
	Valid when one intentionally focuses on the no‑jump branch rather than full steady‑state dynamics.

	For the jump operator $L = |0\rangle\langle1|$ and the feedback operator $F \propto |0\rangle\langle1| + |1\rangle\langle0|$, substituting into Eq.~(\ref{Hnh}) yields only the projector $|1\rangle\langle1|$ in the combinations $L^\dagger L$, $L^\dagger F$ and $F L$, regardless of the feedback strength $G$, so no gain on the ground state is generated and true $\mathcal{PT}$ symmetry cannot be realised. Moreover, after substituting the term $\frac{G\sqrt{\gamma_1}}{2}\,\bigl[L^\dagger F - F\,L\bigr]$ in Hamiltonian~(\ref{Hnh}) contains no imaginary component unless one chooses $F \propto \sigma_y$, and even in that case the $|0\rangle\langle0|$ term remains absent. To address these issues, we introduce an auxiliary state to enable the desired $|1\rangle\langle0|$ transition, as discussed below.

	\subsection{Achieving Gain via Feedback in a $\Lambda$-Type Quantum System}
	
	To achieve gain on $|0\rangle\langle0|$, we introduce an auxiliary level $|a\rangle$ to form a $\Lambda$-type configuration with levels $|0\rangle$, $|1\rangle$, and $|a\rangle$. The transition $|0\rangle \leftrightarrow |a\rangle$ is driven with a Rabi frequency $\Omega_a$ and detuning $\Delta_a$, while the level $|a\rangle$ spontaneously decays to $|1\rangle$ at a rate $\gamma_a$, described by the Lindblad operator $\mathcal{L} = \sqrt{\gamma_a} |1\rangle\langle a|$.
	
	Following the effective operator formalism of Reiter and Sørensen \cite{Reiter2012, ReiterNJP}, we derive the effective jump operator as:
	\begin{equation}
		\mathcal{L}_{\text{eff}} = \mathcal{L} \cdot (H_{\text{NH}})^{-1} \cdot V_+,
	\end{equation}
	where $	H_{\text{NH}} = \Delta_a - i \frac{\gamma_a}{2}$, $
	V_+ = \frac{\Omega}{2} |a\rangle\langle 0|$, and
	$\mathcal{L} = \sqrt{\gamma_a} |1\rangle\langle a|$.
	Substituting these into the expression, we obtain (for $\Delta_a = 0$)
	\begin{equation}
		\mathcal{L}_{\text{eff}} = i \sqrt{\gamma_{\text{eff}}} |1\rangle\langle 0|
	\end{equation}
	by defining $\gamma_{\text{eff}} = \frac{\Omega^2}{\gamma_a}$.
	
	If feedback is applied based on the measurement of this effective decay from $|0\rangle$ to $|1\rangle$ (rather than the natural decay from $|1\rangle$ to $|0\rangle$), the effective non-Hermitian Hamiltonian becomes \cite{eff}
	\begin{equation}\label{feedbackwitheffect}
		H_{\text{eff}} = H_{\text{sys}} - \frac{i}{2} \mathcal{L}_{\text{eff}}^\dagger \mathcal{L}_{\text{eff}} - \frac{G}{2} \bigl[ \mathcal{L}_{\text{eff}}^\dagger F - F \mathcal{L}_{\text{eff}} \bigr],
	\end{equation}
	where $G$ is the feedback gain and $F$ is the feedback operator. Additionally, accounting for the natural decay from $|1\rangle$ to $|0\rangle$ at rate $\gamma_{10}$ (with Lindblad operator $\sqrt{\gamma_{10}} |0\rangle\langle 1|$), the effective Hamiltonian is
	\begin{equation}
		H_{\text{eff}} = H_{\text{sys}} - \frac{i \gamma_{10}}{2} |1\rangle\langle 1| - \frac{i \gamma_{\text{eff}}}{2} |0\rangle\langle 0| + i G \sqrt{\gamma_{\text{eff}}} |0\rangle\langle 0|.
	\end{equation}
	
	To engineer a specific non-Hermitian dynamics, we set the feedback gain $G$ to satisfy
	\begin{equation}\label{Gequal}
		G \sqrt{\gamma_{\text{eff}}} - \frac{\gamma_{\text{eff}}}{2} = \frac{\gamma_{10}}{2}.
	\end{equation}
	Thus, the effective Hamiltonian becomes
	\begin{equation}\label{idealpt}
		H_{\text{eff}} = H_{\text{sys}} - \frac{i \gamma_{10}}{2} |1\rangle\langle 1| + \frac{i \gamma_{10}}{2} |0\rangle\langle 0|.
	\end{equation}
	This form represents a balanced gain-loss non-Hermitian Hamiltonian, with loss on $|1\rangle$ and gain on $|0\rangle$, which is useful for studying phenomena such as parity-time symmetry or exceptional points in quantum systems.
	
	\section{Discussion}\label{sec4}
	
	In this work, we have developed a method to achieve gain on the ground state in an effectice two-level quantum system through WMQF theory. To finally achieve the goal, we introduce an auxiliary excited state and use the SEOM, we derived an effective jump operator that approximates the dynamics induced by the auxiliary level $|a\rangle$. By applying feedback based on this effective decay, we obtained an effective non-Hermitian Hamiltonian with a balanced gain-loss structure, given by $H_{\text{eff}} = H_{\text{sys}} - \frac{i \gamma_{10}}{2} \left( |1\rangle\langle 1| - |0\rangle\langle 0| \right)$. This Hamiltonian can exhibit characteristics similar to $\mathcal{PT}$-symmetric systems, which are known for phenomena such as exceptional points and/or non-reciprocal behavior, making this system a promising platform for studying non-Hermitian quantum mechanics.
	The accuracy of the effective operator formalism depends on the assumption that the auxiliary level $|a\rangle$ remains weakly populated. This condition holds when the Rabi frequency $\Omega_a$ is significantly smaller than the decay rate $\gamma_a$, i.e., $\Omega_a \ll \gamma_a$~\cite{Kastoryano2011,Reiter2012,ReiterNJP}.
	
\begin{figure}[h]
		\centering
		\includegraphics[width=\linewidth]{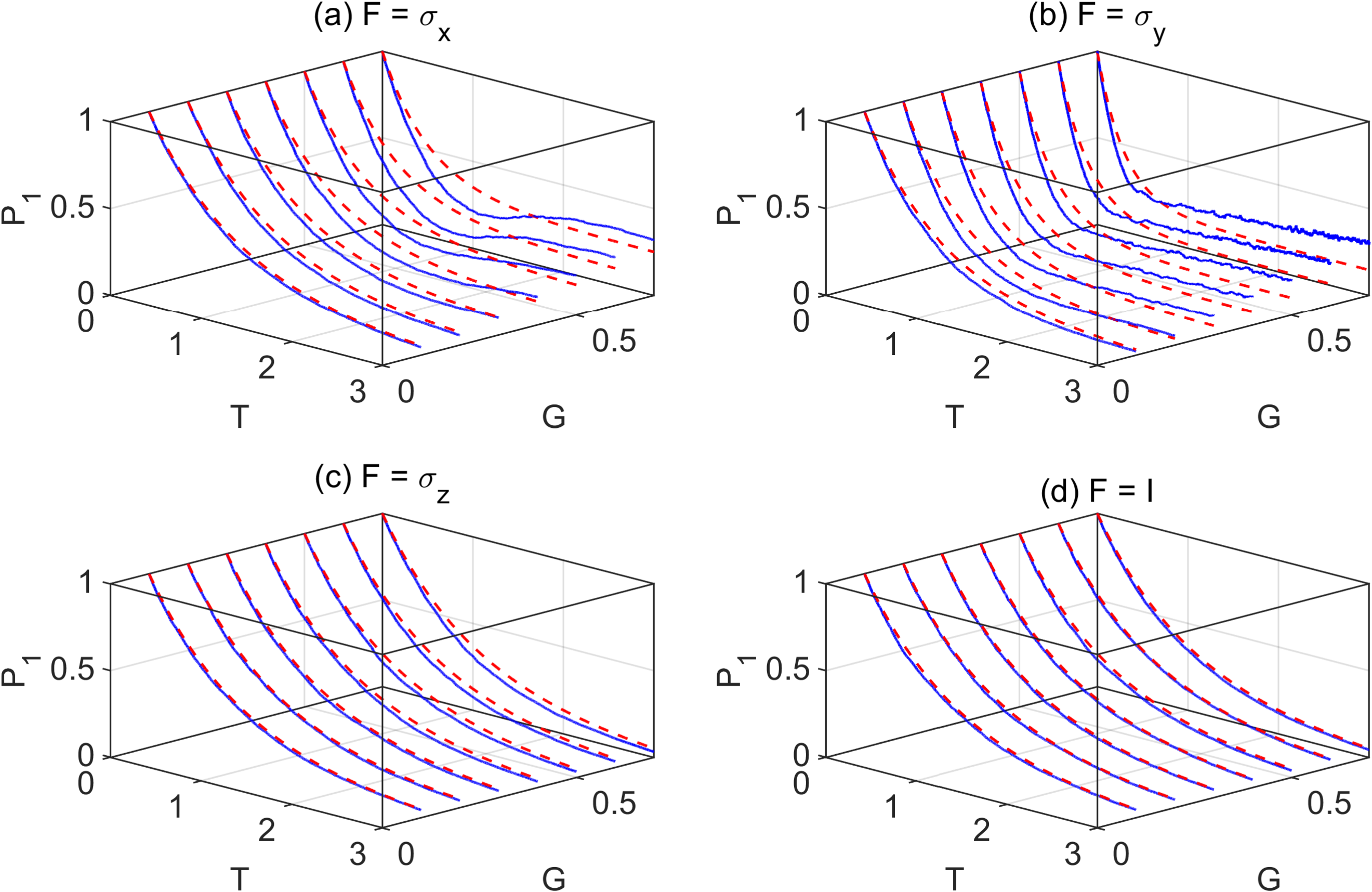}
		\caption{
			Comparison of \(P_{1}\) (population in state $|1\rangle$) as a function of evolution time and feedback strength, computed by (i) averaging the SME‑based feedback master equation \ref{SME} (dashed red line) and (ii) the ideal unconditional feedback master equation \eqref{unconditionalmaster} (solid blue line), for feedback operations\{$\sigma_x,~\sigma_y,~\sigma_z,~ I$\} (a)–(d). Results are shown for \(N_{\mathrm{traj}}=1000\) trajectories, with the system initialized in \(\lvert1\rangle\) and $\gamma_1$ is set as 1 and $L = |0\rangle\langle1|$ in Eqs.~ (\ref{SME}) and (\ref{unconditionalmaster}).}
		\label{fig2}
	\end{figure}

	We have to stress two things for the scheme. Firstly, this is a ensemble average behavior of conditional master equation with feedback. Thus, the effectiveness from Eq.~(\ref{SME}) to Eq.~(\ref{unconditionalmaster}) is important for the following process. In Fig.~\ref{fig2}, we consider $N_{\mathrm{traj}}$ number trajectories to average SME dynamics and compare with the resuts simulated by unconditional master equation. The results show that the approximation is valid for different \emph{G} and feedback operators. Secondly, dropping the jump term is an approximation whenever either (i) post‑select on trajectories with zero detection clicks, or (ii) restrict attention to time intervals so short (or jump rates so low) that the probability of any jump is negligible. Dropping the jump term \(\tilde{c}\rho\,\tilde{c}^\dagger\) isolates the \emph{conditional} (no‑jump) evolution, either via post‑selection on no clicks or in the short‑time/weak‑jump limit \(\langle \tilde{c}^\dagger \tilde{c}\rangle \,\Delta t \ll 1\). It should be note that for case (i) there is no logical error because the the quantum jump based feedback is based on $\mathcal{L}_{\rm eff}$(relevant to $|1\rangle\langle0|$) not based on $\tilde{c}$ that have been discarded the jump process.  
	
	\begin{figure}[h]
		\centering
		\includegraphics[width=8.5cm,height=5 cm]{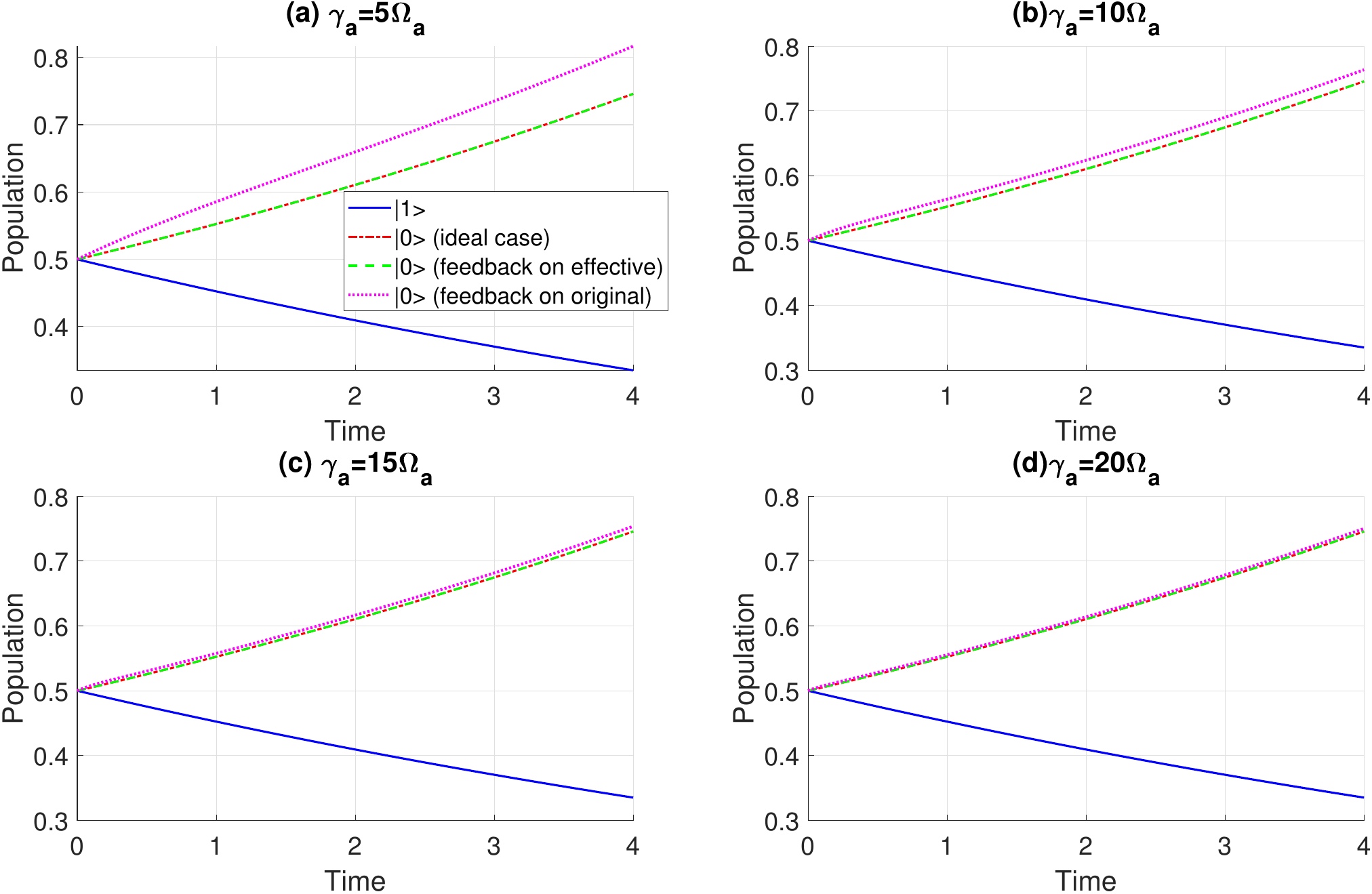}
		\caption{
			Numerically calculated population dynamics versus evolution time for (i) the ideal $\mathcal{PT}$‑symmetric Hamiltonian dynamics, (ii) the $\mathcal{PT}$‑symmetric Hamiltonian dynamics with gain achieved via the effective-operator formalism based quantum feedback and \emph{G} is set through Eq.~(\ref{Gequal}), and with loss the same as case (i), (iii) the $\mathcal{PT}$‑symmetric Hamiltonian dynamics with gain achieved via the original $\Lambda$-type three-level system,  \emph{G} and loss process are the same as (ii). The parameters for subfigures(a)-(d) are $\{\Omega_a,~\gamma_a\} = \{0.5, 2.5\}$, $\{1, 10\}$, $\{1.5, 22.5\}$, $\{2, 40\}$, respectively. $\Delta_a = 0$ induces $\gamma_{\rm eff, 0\rightarrow1}=0.1$ for all cases. The natural decay rate $\gamma_{1\rightarrow0}$ is set as 0.1.
		}
		\label{fig3}
	\end{figure}

	In Fig.~\ref{fig3}, we initialized the system state as $(|0\rangle\langle0|+|1\rangle\langle1|)/2$ and tracked the population evolution versus the evolution time. These simulations were implemented by solving the dynamics governed by the Hamiltonians~(\ref{idealpt}) for ideal case, and  (\ref{feedbackwitheffect}) for feedback on effective operator method, respectively.  And Eq.~(\ref{Hnh}) with decay channel and strength from $\Lambda$-type configuration is used to simulate the original dynamics. Here we should point out that the S{\o}rensen-Reitor effective method is valid for the weak driving condtion $\gamma_a\gg\Omega_a$, and this is consistent with the results in Fig.~\ref{fig3}. One can see from Fig.~\ref{fig3} that for the same $\gamma_{\rm eff}=\Omega_a^2/\gamma_a$, the results get by effective dynamics and that by original dynamics fit well each other when $\Omega_a\ll\gamma_a$ is well satisfied. This close match validates the approximation in the weak driving regime, though deviations occur as $\Omega_a / \gamma_a$ increases due to greater population in $|a\rangle$.
	
	Experimentally, the implementation of this quantum feedback based scheme requires rapid measurement and control, feasible with modern quantum technologies~\cite{wisemanbook,Riste193,Vijay194,Weber407}. However, practical challenges such as finite feedback delay or measurement inefficiencies could affect the dynamics, warranting further investigation in future studies~\cite{Zhang2017}.
	
	It is worth noting that non-Hermitian dynamics are inherently non-unitary. Unlike purely loss processes where the state norm decays due to loss, the presence of gain can cause the norm to grow beyond unity. Such growth requires renormalization to maintain physical interpretability~\cite{djf2019}. The enhancement of state norm in our approach is physically meaningful, as it reflects an increased quantum jump probability, thereby boosting the success rate of postselected measurements.

	\section{Conclusion}\label{sec5}
	In conclusioin, this study shows the power of combining WMQF and SEOM to engineer the gain in quantum systems to achieve truly $\mathcal{PT}$-symmetry Hamiltonian. The controlled introduction of gain offers potential applications in quantum sensing, amplification, and the study of topological phenomena in open systems.
	Numerical simulations demonstrate that the effective dynamics is well captures the system dynamics under various feedback operations, with improved agreement as the weak-driving condition is more precisely satisfied, thereby verifying the validity of our approach. Future work could extend this approach to multi-level systems or incorporate time-dependent controls to further manipulate the gain-loss profile, enhancing its utility in quantum technologies.
	
	\begin{acknowledgements} 
		This work was supported by the National Natural Science Foundation of China (Grants No. 12274376 and No. 12575032), a major science and technology project of Henan Province under Grant No. 221100210400, and the Natural Science Foundation of Henan Province under Grant No. 232300421075.
	\end{acknowledgements}


\begin{thebibliography}{9}
\bibitem{eff}
		Here, we should point out that during the unconditional master equation derivation, the decay rate and decay channel are seperated for derivation, here we together them.
\end{thebibliography}
\end{document}